\documentclass{elsart}
\usepackage{amstex,graphicx}
\DeclareMathAlphabet\cal{U}{eus}{m}{n}

\begin{document}

\begin{frontmatter}
\title{Bulk properties of rotating nuclei and the validity of the liquid drop
model at finite angular momenta \thanksref{support}}
\thanks[support]{Work supported in part through funds granted by a
French-Algerian agreement between CNRS and DRS, by a French-Bulgarian
agreement between CNRS and the Academy of Sciences of Bulgaria and by the
Bulgarian National Scientific foundation under contract F621.}
\author[sofia]{J. Piperova,}
\author[cenbg]{D. Samsoen,}
\author[cenbg]{P. Quentin,}
\author[cenbg,setif]{K. Bencheikh,}
\author[ires]{J. Bartel}
\author[ipn]{and J. Meyer}
\address[sofia]{Institute of Nuclear Research and Nuclear Energy (Bulgarian
Academy of Sciences), Tzarigradsko Chaussee 72, 1784 Sofia, Bulgaria}
\address[cenbg]{Centre d'Etudes Nucl\'eaires de Bordeaux-Gradignan (IN2P3-CNRS
and Universit\'e Bordeaux-1), BP 120, 33175 Gradignan-Cedex, France}
\address[setif]{D\'epartement de Physique, Universit\'e de S\'etif, S\'etif,
Algeria}
\address[ires]{Institut de Recherches Subatomique (IN2P3-CNRS and Universit\'e
Louis Pasteur), BP20, 67037 Strasbourg-Cedex, France}
\address[ipn]{Institut de Physique Nucl\'eaire de Lyon (IN2P3-CNRS and
Universit\'e Claude Bernard), 43 Boulevard du 11 Novembre 1918, 69622
Villeurbanne-Cedex, France}
\maketitle

\begin{abstract}
Out of self-consistent semi-classical calculations performed within the
so-called Extended Thomas-Fermi approach for 212 nuclei at all even angular
momentum values $I$ ranging between 0 and 80 $\hbar$ and using the Skyrme
SkM$^{*}$ effective force, the $I$-dependence of associated liquid drop model
parameters has been studied. The latter have been obtained trough separate
fits of the calculated values of the strong interaction as well as direct and
exchange Coulomb energies. The theoretical data basis so obtained, has allowed
to make a rough quantitative assessment of the variation with I of the usual
volume and surface energy parameters up to spin of $\sim30$-40 $\hbar$. As a
result of the combined variation of the surface and Coulomb energies, it has
been shown that this $I$-dependence results in a significant enhancement of
the fission stability of very heavy nuclei, balancing thus partially the
well-known instability due to centrifugal forces.
\end{abstract}
\end{frontmatter}

\section{Introduction}

The availability of heavy ion beams at sufficiently high energy has made it
possible to transfer very high angular momenta to nuclei, particularly in
fusion-evaporation channels. Upon increasing the angular momentum one has
opened up the exploration of a wide range of very large nuclear deformations,
as in superdeformed states which were previously reached only in the fission
process of ground (or weakly excited) states of heavy nuclei.  As a matter of
fact, one is currently able to reach the critical angular velocity regimes
where the nuclear system experiences a centrifugal disassembly on all the
accessible part of the nuclear chart. This is why a long time ago already,
various authors have studied the equilibrium shapes and stability properties
of rotating nuclei within a liquid drop model approach, drawing a close
analogy between the properties of such a self-bound mesoscopic system with
those of self-gravitating and rotating celestial objects (see Ref. \cite{CPS82}
and Refs. quoted therein). Later, a more microscopic description of such
nuclear systems has been provided upon minimizing the so-called ``Routhian''
variational quantity in a macroscopic-microscopic approach (from the
pioneering works of Refs. \cite{NPF76,ALL76} to more systematic approaches as
e.g. in Ref. \cite{JD}) thus relying again on the liquid drop model assumption
when using in practice the Strutinsky energy expansion \cite{S67} in a
slightly, and easily, transposed version. More recently, fully self-consistent
Routhian approaches (after preliminary attempts making use of an inert core
approximation \cite{BMR73,G76}), have been made available either with the
Skyrme \cite{PM76,BFH87,DD95,SQ98} or the Gogny \cite{ER93,GD94}
parameterization of the nucleon-nucleon effective interaction and have been 
shown to be reasonably realistic.

While the latter type of calculations does not imply any liquid drop model
assumption to study the behavior of rotating nuclei, they are not really easy
to handle in systematic calculations and certainly not suited, due to shell
effects, to provide a transparent account of the bulk properties of nuclei at
finite angular momenta. Yet, one would like to take stock on the impressive
ability of such effective force parameterizations (Skyrme SkM$^{*}$ \cite{BQ82}
or SLy4 \cite{JM} and Gogny D1S \cite{JD1} interactions) to yield a good
reproduction of a wide array of static and (low excitation energy) dynamical
properties. As well known, the theoretical framework to search for the bulk
nuclear properties embedded in such purely microscopic approaches is a
semi-classical approach \`{a} la Wigner \cite{W32}, i.e. performing a truncated
expansion of the solution (namely of the one-body reduced density matrix in
the Hartree-Fock case). More specifically, we will use here the Extended
Thomas-Fermi approach proposed in Ref. \cite{BR71} and exhaustively discussed
in Ref. \cite{BGH85}.

In our Routhian approach we make use of a Skyrme type effective force
parameterization. We explicitly express the functional dependence of some local
density functions (as e.g. current and spin-vector densities) providing the
relevant informations on the non-local (in $\boldsymbol{r}$) part of the
time-odd component of the one-body density matrix, in terms of the spin-scalar
density $\rho(\boldsymbol{r})$. Some steps in that direction have been made by
Grammaticos and Voros both with \cite{GV80} and without \cite{GV79} spin
degrees of freedom. However the full analytical solution of this problem has
only been provided recently \cite{BQ94} and used to perform self-consistent
semi-classical calculations \cite{CM94}. The present description of bulk
nuclear properties at finite angular momentum is clearly an extension of these
previous studies. It fully develops some ideas presented earlier \cite{CM94}
in a somewhat exploratory fashion.

Indeed, we want to assess the relevance of the concept of a liquid drop for
the description of rotating nuclei at equilibrium. The rationale for doing it
stems from the anisotropic character of the rotational constraint imposed on
the considered piece of nuclear matter which seems at first sight to be at
variance with the isotropic character attached to a saturated liquid
drop. Furthermore nuclei being leptodermous, as nicely illustrated by Myers
and Swiatecki \cite{MS69}, one may wonder whether or not the surface tension,
and hence the skin properties, might not be affected by the rotational
character of the nuclear fluid. The above may be summarized in a practical
form by recalling first that in Ref. \cite{CPS82}, for instance, the nuclear
energy is written, with usual notation, as
\begin{equation}
E(\beta,I) = E(\beta,I=0) + \frac{(I\hbar)^{2}}{2 \Im^{(2)}} ,
\label{ldm-energy}
\end{equation}
where $\beta $ stands for a set of deformation parameters. One takes now for
the energy $E(\beta,I=0)$ a standard version of the liquid drop model and the
rigid body moment of inertia associated with the given shape for the parameter
$\Im^{(2)}$. Therefore one assumes that all the $I$-dependence is only 
contained in the kinetic collective energy of a rigid rotation.

Let us translate that into the Hartree-Fock language. One may then split the
total energy in the laboratory frame into two pieces:
\begin{equation}
E = E_{+} + E_{-} ,
\label{energy-split}
\end{equation}
the first involving only the part $\rho_{+}$ of the density matrix which is
even with respect to the time-reversal symmetry and the second, $E_{-}$,
issuing from the time-odd part $\rho _{-}$ of the density matrix (despite the
two-body character of the interaction part of the Hamiltonian $H$, there are
no cross-terms in Eq. (\ref{energy-split}) due to the time reversal invariance
of $H $). For the standard Skyrme energy density functionals, the former part
involves local densities (for both charge states) like the spin-scalar
diagonal densities, the kinetic energy densities and the so-called spin-orbit
densities (see e.g.  Ref. \cite{VB72}), while the latter involves the current
densities and the spin-vector diagonal densities (see
e.g. Ref. \cite{BFH87}). In the adiabatic limit, such a partition into two
energies does correspond to the partition assumed in
Eq. (\ref{ldm-energy}). In non-adiabatic regimes, the dependence of the two
parts $\rho_{+}$ and $\rho_{-}$ of $\rho $ in terms of the collective velocity
(in the present case the angular velocity $\omega $) is more complicated since
$\rho_+$ may contain, for instance, a quadratic dependence on $\omega $.

One is thus lead, in practice to the two following questions:

-- Is it correct that in a self-consistent semi-classical approach $E_{-}$ may
be written as in Eq. (\ref{ldm-energy}) with a rigid body moment of inertia? 
\\
-- Is it correct to assume that $E_{+}$ may be well approximated by a liquid
drop energy, and if so are the parameters of this liquid drop independent on
the angular momentum?

The answer to the first question is essentially affirmative as it has been
demonstrated in Ref. \cite{BQ94}. There, it has been shown that the
semi-classical moment of inertia $\Im$ may be split into two parts for which
compact mathematical expressions have been given in the Skyrme force case.
The first one is the zero order term of the semi-classical expansion (the
so-called Thomas-Fermi term) and is as expected (and well known before the
work of Ref. \cite{BQ94}) the rigid body moment of inertia. The remaining part
corresponds to the second order term of this expansion which can be decomposed
into two contributions, one coming from the orbital, the other from the spin
degrees of freedom. They may be illustrated upon using a well established
analogy between an angular velocity and a magnetic field $\boldsymbol{B}$
(analogy between, for instance, the Coriolis pseudo-force and the Lorentz
magnetic force, first advocated in the context of a rotating many fermion
system by Dabrowski \cite{D75}). For realistic effective forces (in
Ref. \cite{BQ94}, we have used the SkM$^{*}$ force which is known to give, at
least in the $\beta $-stability valley a very satisfactory account of bulk
nuclear properties) one has found a negative contribution of the orbital
degrees of freedom to $\Im$, corresponding to an analog of a Landau
diamagnetism, while the spin degrees of freedom contribution being positive,
corresponds to a Pauli paramagnetic alignment. Both contributions are rather
weak with respect to the Thomas-Fermi contribution, and moreover, as we have
seen, they cancel each other to a large extent. Specifically, it has been
found in Ref. \cite{BQ94} that one can write with obvious notation for a
nucleus of mass $A$
\begin{equation}
\Im = \Im_{RB} [1 + (\eta_{l} +\eta_{s}) A^{-2/3}]
\end{equation}
with the following values for the coefficients of the second order
contributions (again for the SkM$^{*}$ force):
\begin{equation}
\eta_{l} \approx -1.94 , \hspace{5mm} \eta_{s} \approx 2.63 ,
\end{equation}
leaving thus only a small correction to the rigid body ansatz, especially
for heavy nuclei.

The problem of the kinetic collective energy being settled, the second
question raised above is still open. First, is there any $I$-dependence in the
non-kinetic ($E_{+}$) part of the energy? The answer to this is clearly
yes. In Ref. \cite{CM94}, we have found that the self-consistent semi-classical
energy $E_{+}$, for a given nucleus and using the SkM$^{*}$ force, was indeed
very much varying. Therefore, if at all valid, the liquid drop model to be
used should clearly incorporate coefficients varying with the rotational
velocity. An important question needs to be clarified at this point. A priori,
there are no clear reasons why the liquid drop parameters should vary with $I$
or $\omega $ . Of course, the former being an observable which is furthermore a
conserved quantity, seems a better candidate. But it is not clear to us why
this quantal feature should reflect itself in this way for the rotational
dependence of a rather crude modelization of the energy. However in
Ref. \cite{CM94}, it has been shown that the variation of the energies $E_{+}$
were of similar relative magnitude between light and heavy nuclei, only when
compared for the same value of $I$ (and not for the same value of $\omega $
which scales as $A^{-5/3}$ for a given value of $I$). We are thus lead to
conclude phenomenologically that, if valid, the rotating liquid drop model
should be $I$-dependent and not $\omega $-dependent. (Note that even though,
as stated above, the results of Ref. \cite{CM94} may lead to such a
conclusion, the liquid drop model parameter fit in terms of the angular
velocity performed there is somehow inconsistent).

The remaining issue is then the validity of the liquid drop model at finite
angular momenta. For that purpose, we need to perform for each value of the
angular momentum which we decide to study, high accuracy self-consistent
semi-classical calculations for a sufficient number of nuclei covering
reasonably well the chart of nuclides so that we should be able to make a
significant liquid drop parameter fit. This was not at all the case of the
calculations of Ref. \cite{CM94} (too few nuclei, 33, and an accuracy in the
variational semi-classical solutions somewhat insufficient for the present
purpose). It is the aim of the present paper therefore to achieve in a
numerically satisfactory fashion the preliminary calculations sketched in
Ref. \cite{CM94}.

Two remarks are worthwhile at this point. First, it is clear that we do not
include pairing correlations. This could be done, as we are currently
attempting, within the BCS or the Hartree-Fock-Bogoliubov framework in the
semi-classical approximation (for the Thomas-Fermi and Extended Thomas-Fermi
non-rotating case see Refs. \cite{BS80} and \cite{S1} respectively). The
second remark deals with the bulk deformation properties. All our
self-consistent semi-classical calculations will be performed at sphericity.
As a matter of fact when assessing the validity of a liquid-drop model
approximation, one intends to do two different things. The first is to check
the ability of a liquid drop parameterization to reproduce with a good accuracy
the semi-classical energies at a given deformation. For that purpose choosing
spherical solutions is an obvious choice for the sake of computational
convenience. The second is to check whether or not the semi-classical energies
are varying with deformation like a sum of various geometrical quantities
associated with the deformation (surface, Coulomb energy or more). This will
not be attempted here. Yet, we think obvious that the present
approach casts some light on nuclear deformation properties at finite angular
momentum (through the fissility parameter for instance). Restricting ourselves
to spherical shapes, we may therefore ignore the difficulties described by
Krappe, Nix and Sierk \cite{KN}, and studied in the rotational context in
Refs. \cite{S,mumus}, when dealing with nuclear shapes having some concavity
(like fission saddle points of some nuclei) or not simply connected (like
post scission or fusion configurations, for instance).

The paper will be organized as follows. Section 2 will detail the theoretical
framework in which we have studied the bulk rotational properties of the
nuclei. In Section 3 we will present the calculational approach insisting on
the choice of the sample of nuclei, the accuracy of the self-consistent
calculations and the fitting procedure. The results will be displayed and
discussed in Section 4. Finally, Section 5 will contain some conclusions and
perspectives of the present work.

\section{The theoretical framework}

To describe in a semi-quantal way the wave function of a nucleus rotating
with respect to the laboratory frame with an angular velocity $\omega $, one
minimizes the Routhian $R$ defined in terms of the microscopic Hamiltonian $H$
as
\begin{equation}
R = H - \omega I .
\label{routhian}
\end{equation}
In the Hartree-Fock approximation, the single particle states defining the
solution is thus reduced to an eigenvalue problem involving the one-body
reduction of the Routhian of Eq. (\ref{routhian}). The corresponding
semi-classical one-body density matrix may thus be obtained according to the
usual Extended Thomas-Fermi method upon replacing the Hartree-Fock one-body
Hamiltonian by the corresponding one-body Routhian. For a Skyrme effective
force, the Routhian expectation value is obtained as an integral over space of
a Routhian density which is expressed (see e.g. Ref. \cite{BFH87}) in terms of
various densities (spin-scalar and spin-vector densities, kinetic energy
densities, spin-orbit and current densities). In the semi-classical Extended
Thomas-Fermi approach all densities may be expressed as functionals of the
spin-scalar densities \cite{BQ94}. Therefore the Routhian density is itself a
functional of the spin-scalar densities. The variational solution for a given
angular velocity may be just obtained then by minimizing the integral of this
Routhian functional.

To do so, choosing the $Oz$ axis as the rotation axis, we have used a
parameterization of the density functions $\rho_q(\boldsymbol{r})$ (here
called $\rho_q^\mathrm{new}$ for reasons explained below), where $q$
stands for the charge state, expressed in cylindric coordinates 
$(r,z,\varphi )$ as
\begin{eqnarray}
\rho_q^\mathrm{new}(r,z) = \rho_{0q} &&\left[1 - \alpha_q \exp(-\beta_q^2 r^2 
s_q^{2/3})\right] \nonumber \\
&& \left[1 + \exp \left(\frac{\sqrt{r^2 s_q^{2/3} + z^2 s_q^{-4/3}} -
R_q}{\sigma_q} \right)\right] ^{-\gamma_q} .
\label{anisotropic}
\end{eqnarray}
Whereas we want to describe spherical solutions, they appear a priori to
have only a cylindrical symmetry (see however the discussion below). The
densities depend on 14 parameters (7 for each charge state). However the
space integrals of the densities $\rho_q(\boldsymbol{r})$ should be equal to
$N_q$, the particle number of charge $q$, leaving thus 12 independent
parameters.  For $\alpha _q=0$, and $s_q=1$, one is left with surface
asymmetric (for $\gamma _q\neq 1$) Fermi distributions as widely used in
Ref. \cite{BGH85} for instance
\begin{equation}
\rho_{q}(\boldsymbol{r}) = \rho_{0q} \left[1 + \exp \left(\frac{\sqrt{x^{2}
+ y^{2} + z^{2}} - R_{q}}{\sigma_{q}} \right)\right] ^{-\gamma_{q}} .
\label{isotropic}
\end{equation}
One notices in Eq. (\ref{anisotropic}) that in front of the modified Fermi
term we have enclosed a factor similar to what has been considered in
Ref. \cite{BGH85} with a very significant difference, namely its anisotropic
character. This has been introduced to let the solution adjust to the
centrifugal forces which are themselves not isotropic. For the same reasons
advocated in the slightly different context of Ref. \cite{BGH85}, we have
reduced the variational space by imposing that $\beta _q$ should have a fixed
value within a range where this modification would not alter the desired
leptodermous and smooth character of the semi-classical densities. In practice
we have used:
\begin{equation}
\beta_q^2 = 1/(c_q R_q)^2,\hspace{5mm} c_q = 0.5 ,
\end{equation}
$R_q$ being the radius parameter introduced in Eq. (\ref{anisotropic})

Let us discuss now the scaling parameter $s_q$ appearing in
Eq. (\ref{anisotropic}).  Introducing an anisotropy as above described, our
solution lacks its spherically symmetrical character. To be able to compare
solutions with various values of the parameters $\alpha _q$ we must constrain
them to keep, in some way, a vanishing deformation. This is achieved by
imposing that
\begin{equation}
\langle r^2\rangle_q = 2 \langle z^2\rangle_q,
\end{equation}
where $\langle r^2\rangle_q$ and $\langle z^2\rangle_q$ stand for the
expectation values of $r^2$ and $z^2$ for the distribution of nucleons of
charge $q$. Taking into account the particle number conservation, one gets the
following explicit expression for the parameter $s_q$:
\begin{equation}
s_q^{-2} = 2\int_0^\infty r \;\mathrm{d}r \int_{-\infty }^\infty z^2
\rho_q^\mathrm{old}\; \mathrm{d}z \;\left[ \int_0^\infty r^3 \;\mathrm{d}r
\int_{-\infty }^\infty \rho_q^\mathrm{old}\; \mathrm{d}z \right]^{-1},
\end{equation}
with
\begin{equation}
\rho _q^\mathrm{old}(r,z) = \rho _{0q} \left[1 - \alpha_q \exp(-\beta_q^2
r^2)\right] \left[ 1 + \exp \left(\frac{\sqrt{r^2+z^2}-R_q}{\sigma_q}\right)
\right]^{-\gamma_q} .
\end{equation}
To obtain the expression for $s_q^2$ one may introduce stretched coordinates 
$\tilde r=s_q^{1/3}r$ and $\tilde z=s_q^{-2/3}z$ in terms of which one has 
$\rho_q^\mathrm{new}(\tilde r,\tilde z)=\rho _q^\mathrm{old}(r,z)$.

The usual iterative procedure now begins with the determination of the
scaling factor $s_q.$ Then we find the parameter $\rho_{0q}$ from the
normalization condition
\begin{equation}
N_q = 4\pi \int_0^{R\max } r \;\mathrm{d}r \int_0^{z_{\max }} 
\rho_q^\mathrm{new}(r,z) \;\mathrm{d}z ,
\end{equation}
Hereafter $N_p=Z,$ $N_n=N$. In all the integrals we take advantage of the
reflexion symmetry $\rho_q^\mathrm{new}(r,-z)=\rho_q^\mathrm{new}(r,z)$ .

Once the variational parameters have been determined, one obtains the total
energy in the laboratory frame according to the decomposition of
Eq. (\ref{energy-split}) where $E_{+}$ is obtained by using the time-even
energy Skyrme functional of Ref. \cite{VB72}. One gets namely (see
Ref. \cite{BQ94}).
\begin{equation}
E[\rho] = \int {\cal H}[\rho] \;\mathrm{d}^3r =\int {\cal H}_\mathrm{stat}
[\rho] \;\mathrm{d}^3r + \frac{1}{2}\Im^{(2)}(\rho)\omega ^2 ,
\end{equation}
with
\begin{equation}
{\cal H}_\mathrm{stat}[\rho] = \frac{\hbar^2}{2m} \sum_q f_q \tau_q(\rho) + 
V_q^\mathrm{central}(\rho) + V_q^\mathrm{s.o.}(\rho) ,
\label{H-stat}
\end{equation}
where $f_q$ is the usual Skyrme effective mass factor and $\tau _q(\rho)$ the
ETF expression for the static kinetic energy density up to $\hbar ^4$-terms
(see e.g. Ref. \cite{BGH85}). The expression for $V_q^\mathrm{central}(\rho)$
is given with the notation of Ref. \cite{BFH87} for the Skyrme parameters
($B_{i},\alpha $) where as a general rule all densities without a charge state
subscript $q$ are total densities (i.e. for instance $\rho =\sum_q\rho _q$).
\begin{eqnarray}
V_q^\mathrm{central} &=& B_1\rho ^2 +B_2\sum_q\rho_q^2 +B_5\rho\nabla^2\rho
+ B_6\sum_q\rho_q\nabla^2\rho_q \nonumber \\
&&+ \big(B_7\rho^2 + B_8\sum_q\rho_q^2\big) \rho^\alpha
\end{eqnarray}
The ETF expression for the spin-orbit contribution up to $\hbar ^4$-terms
$V_q^\mathrm{s.o.}$ to ${\cal H}_\mathrm{stat}(\rho )$ is given e.g. in
Eq. (A5) of the review paper by Brack et al. \cite{BGH85}. Finally, the ETF
expression of $\Im^{(2)}$ up to $\hbar^2$-terms has been derived
in Ref. \cite{BQ94} (see e.g. Eq. (63) there).

To obtain solutions corresponding to precise values of the angular momentum
$I$, we adjust iteratively the value of $\omega $ accordingly. We have made
calculations for a sample of 212 even-even nuclei, covering all the chart of
nuclides whose mass is higher than 34 within reasonable estimates of the two
drip lines and including five long isotopic series (Ca, Ge, Sn, Sm, Pb), as
detailed in Table \ref{liste}. Obviously for semi-classical solutions whose
static properties are smooth functions of $N$ and $Z$, the particular choice
of nuclei is irrelevant provided that the sampling is dense enough in the
region of interest for the properties under study. For instance to get
reasonable estimates of proton-neutron asymmetry properties it is essential to
include in the sample isotopes up to the drip lines. We will discuss in the
next Section the quality of our sample in this respect.

\begin{table}[t]
\begin{center}
\caption{Isotopes used in the fit. For each elements we have considered all
even-A isotopes from $A_i$ to $A_f$ (including the boundaries) corresponding
to a number of isotopes $N_\mathrm{is}$.}
\label{liste}
\bigskip
\begin{tabular}{c|lll|l l c|lll|l}
Element & $Z$ & $A_i$ & $A_f$ & $N_\mathrm{is}$ &$\qquad$&
Element & $Z$ & $A_i$ & $A_f$ & $N_\mathrm{is}$ \\ \cline{1-5}\cline{7-11}
Ca       & 20  &   34  &  78   &  23 && 
Yb       & 70  &  166  &  180  &   8 \\ 
Ni       & 28  &   56  &       &   1 && 
Os       & 76  &  184  &  198  &   8 \\
Ge       & 32  &   58  &  118  &  31 && 
Pb       & 82  &  182  &  242  &  31 \\ 
Zr       & 40  &   86  &  100  &   8 && 
Th       & 90  &  226  &  240  &   8 \\ 
Sn       & 50  &   94  &  168  &  38 && 
Pu       & 94  &  236  &  252  &   9 \\ 
Ce       & 58  &  134  &  148  &   8 && 
Cf       & 98  &  252  &       &   1 \\ \cline{7-11}
Sm       & 62  &  120  &  194  &  38 \\ \cline{1-5}
\end{tabular}
\bigskip
\end{center}
\end{table}

Now, for each value of $I$, we have made a liquid drop model fit of the 212
nuclear energies $E_\mathrm{nuc}$, defined in terms of the energy $E_{+}$ as
\begin{equation}
E_\mathrm{nuc}=E_{+}-E_\mathrm{coul},
\end{equation}
where $E_\mathrm{coul}$ is the total Coulomb energy. The latter includes the
direct term ($E_\mathrm{CD}$) and the so-called ``Slater approximation''
(namely a local density approximation \cite{G52} of the Bethe-Bacher
\cite{BB36} nuclear matter estimate) for the exchange term ($E_\mathrm{CE}$)
whose accuracy has been checked earlier \cite{TQ74}. Both Coulomb energy terms
have been fitted separately (i.e. independently of each other and of the
nuclear energy). We leave for another publication \cite{SBQ} an in-depth
discussion of the liquid drop fit of the Coulomb energies, in particular its
isospin-dependence which is generally overlooked (see a preliminary account of
such a work in the review of Ref. \cite{QS}). Namely, the liquid drop
expression for $E_\mathrm{nuc}$ which we have considered here, is given for
the nucleus labeled by j ($j=1,\cdots N_A$, where $N_A$ is the number of
nuclei entering the fit) as
\begin{eqnarray}
E_\mathrm{nuc}^j(I) &=& a_v(I) \; \Big[1 - k_v(I) (1 - 2Z_j/A_j)^2\Big] \; A_j
\nonumber \\
&& + a_s(I) \; \Big[1 - k_s(I) (1 - 2Z_j/A_j)^2\Big] \; A_j^{2/3} .
\end{eqnarray}

The Coulomb part is the sum of a direct ($E_\mathrm{CD}$) and exchange
($E_\mathrm{CE}$) part whose absolute value is known to be one order of
magnitude smaller than $E_\mathrm{CD}$ (see e.g. Ref. \cite{TQ74}). For the
direct term we have made two fits one which is rather usual (sharp density term
plus a leptodermous surface correction as in Ref. \cite{MS66})
\begin{equation}
E_\mathrm{CD}^j(I) = c(I) \; Z_j^2/A_j^{1/3} - \gamma(I) \; Z_j^2/A_j , 
\label{Ecdir1}
\end{equation}
and another extended one discussed in a forthcoming paper \cite{SBQ} which
takes into account both the surface diffuseness and $(N-Z)$ isotopic effects
\begin{eqnarray}
E_\mathrm{CD}^j(I) &=& c(I) \; \Big[ 1 + \alpha_1(I) \big(1 -
2Z_j/A_j\big)A_j^{-1/3} + \alpha_2(I)A_j^{-1/3} \Big] \; Z_j^2/A_j^{1/3}
\nonumber \\
&&-\gamma (I) \; \Big[ 1 + \beta(I) \big(1 - 2Z_j/A_j\big) \Big] \; Z_j^2/A_j .
\label{Ecdir2}
\end{eqnarray}
For the exchange term ($E_\mathrm{CE}$) we consider the usual exchange term
deduced from an infinite nuclear matter non-diagonal density matrix
\cite{BB36} 
\begin{equation}
E_\mathrm{CE}^j(I) = c'(I) \; Z_j^{4/3}/A_j^{1/3} .
\label{Ecexch}
\end{equation}

\section{Some computational details}

We use the SkM$^{*}$ force with the original set of parameters previously
determined by fitting HF results to experimental data. The fit of the
effective interaction has been made correcting approximately the HF energy by
subtracting only the one-body part of the mean energy of the center of mass
(c.m.) motion, $E_\mathrm{cm}^{(1)}$ (see Sect. 2.3 of \cite{BFQ75}). To be
consistent, in our ETF calculations we will also only take into account that
part of the c.m. energy correction. The introduction of this correction leads
to the replacement of the nucleon mass $m$ by $mA/(A-1)$ in the total kinetic
energy and so in our non-local case we have to replace $\sum_qf_q\tau _q(\rho
)$ by $\sum_q f_q(1-1/A)\tau _q(\rho )$ in the energy density ${\cal H}[\rho
]$ [see Eq. (\ref{H-stat})].

The calculation of the Coulomb energy is the most time consuming part of the
problem due to its range. We take into account both the direct and exchange
parts of this energy:
\begin{equation}
E_\mathrm{Coul} = \frac{e}{2} \int \mathrm{d}^3r_1 \;
\rho_p(\boldsymbol{r}_1)V_C(\boldsymbol{r}_1)+E_\mathrm{Coul}^\mathrm{exc} ,
\end{equation}
with
\begin{equation}
V_C(\boldsymbol{r}_1) = e \int \mathrm{d}^3r_2 \; \frac{ 
\rho_p(\boldsymbol{r}_2)}{\left| \boldsymbol{r}_1 - \boldsymbol{r}_2 \right|} .
\end{equation}
The Coulomb exchange part is treated as already mentioned in the so-called
``Slater approximation'' \cite{G52,BB36,TQ74}:
\begin{equation}
E_\mathrm{Coul}^\mathrm{exc} = -\frac{3e^2}{4} (3/\pi)^{1/3} \int
\mathrm{d}^3r \; \rho_p^{4/3}(\boldsymbol{r}).
\end{equation}
The integrand in the direct part of the Coulomb energy has a logarithmic
singularity at the point $\boldsymbol{r}_{1}=\boldsymbol{r}_{2}$. Therefore
such a formula is not suitable for numerical integration. A way to bypass the
latter difficulty is to use the relation
\begin{equation}
\nabla_{\boldsymbol{r}_2}^{2} \left|\boldsymbol{r}_1 - \boldsymbol{r}_2\right|
= \frac{2}{\left|\boldsymbol{r}_{1} - \boldsymbol{r}_{2}\right|}
\end{equation}
and to carry out two integrations by parts to get:
\begin{equation}
V_C(\boldsymbol{r}_1) = \frac{e}{2} \int \mathrm{d}^3r_2 \; \left| 
\boldsymbol{r}_1 - \boldsymbol{r}_2 \right| \nabla_{\boldsymbol{r}_2}^2 
\rho_p(\boldsymbol{r}_2) .
\end{equation}
With our parameterization (\ref{anisotropic}) of the density it is most
convenient to treat the problem in cylindrical coordinates choosing the axis
of symmetry to be the $Oz$ axis. Integrating over the azimuthal angle $\varphi
_2$ the expression for $V_C(r_1,z_1)$ reduces to \cite{V73}
\begin{eqnarray}
V_C(r_1,z_1) = 2e \int_0^\infty r_2 \; \mathrm{d}r_2 \int_{-\infty }^\infty
&&\mathrm{d}z_2 \; \sqrt{(z_1-z_2)^2+(r_1+r_2)^2} \nonumber\\
&&E(k) \nabla ^2\rho_p(r_2,z_2) ,
\end{eqnarray}
with
\begin{equation}
k^2 = 4 r_1r_2 \big[(z_1-z_2)^2+(r_1+r_2)^2\big]^{-1} 
\end{equation}
and where $E(k)$ denotes the complete elliptic integral of the second kind:
\begin{equation}
E(k) = \int_0^{\pi/2} \sqrt{1-k\sin^2 u} \; \mathrm{d}u .
\end{equation}
We have therefore for the direct part of the Coulomb energy:
\begin{equation}
E_\mathrm{Coul}^\mathrm{dir}= \pi e \int_0^\infty r_1\; \mathrm{d}r_1
\int_{-\infty }^\infty \mathrm{d}z_1 \; \rho_p(r_1,z_1) V_C(r_1,z_1) .
\label{Ec-dir}
\end{equation}
In practice, the infinite limit of integration is replaced by $R=25$ fm, and
we get
\begin{eqnarray}
E_\mathrm{Coul}^\mathrm{dir} = 4\pi e^2 \rho_{0p}^2 \int_0^R && \mathrm{d}r_1 
\int_0^R \mathrm{d}z_1 \int_0^R \mathrm{d}r_2 \int_0^R \mathrm{d}z_2 \; 
 \nonumber\\
&& \big[ f(r_1,z_1,r_2,z_2)+ f(r_1,z_1,r_2,-z_2) \big] .
\end{eqnarray}
This 4-fold integral is evaluated numerically using the routine \textsc{d01fcf}
of the NAG library. The same code has been used for the computation of the
integral expressing the number conservation condition. In the simpler case of
$\alpha _q=0$ and $s_q=1$, i.e. for the parameterization (\ref{isotropic}) of
the densities, a Gauss-Legendre quadrature formula has been used in both
integrals (where the integration interval has been divided in 4 equal pieces
in which 16 integration points have been considered).

In calculating the rest of the energy we use a trapezoidal rule with a step of
0.05 fm. For that part, the integration both in $r$ and $z$ coordinates goes
up to $R=24$ fm. Different tests on the precision of the ETF calculations as
connected to the used procedures for numerical evaluation of the integrals
have been made, of which we will merely give here some
examples. In particular, to illustrate the quality of our estimation of the
direct Coulomb energy as given by equation (\ref{Ec-dir}), for the case of the
heavy \nuc{208}{Pb} nucleus, we may mention that upon computing it with the
same mesh size for $R=25$ fm and $R=50$ fm we obtain the same value up to
seven digits (namely up to better than 100 eV). The relative accuracy for the
calculation of $E_\mathrm{nuc}$ is of the order of 10$^{-6}$. For instance
upon increasing the mesh size by 25 percent and using the standard above
described cut-off, the nuclear energy of \nuc{208}{Pb} varies relatively by
less than 10$^{-7}$. Conversely, for the mesh size in use in our calculations
when increasing the cut-off value of $R=24$ fm by 25 percent, the
corresponding variation for the same \nuc{208}{Pb} nucleus is again of the
order of 10$^{-7}$. For the lightest nuclei under consideration here it is one
order of magnitude larger (but still better than 1 keV in absolute terms).

\begin{figure}[htb]
\centerline{\includegraphics*[width=.8\textwidth]{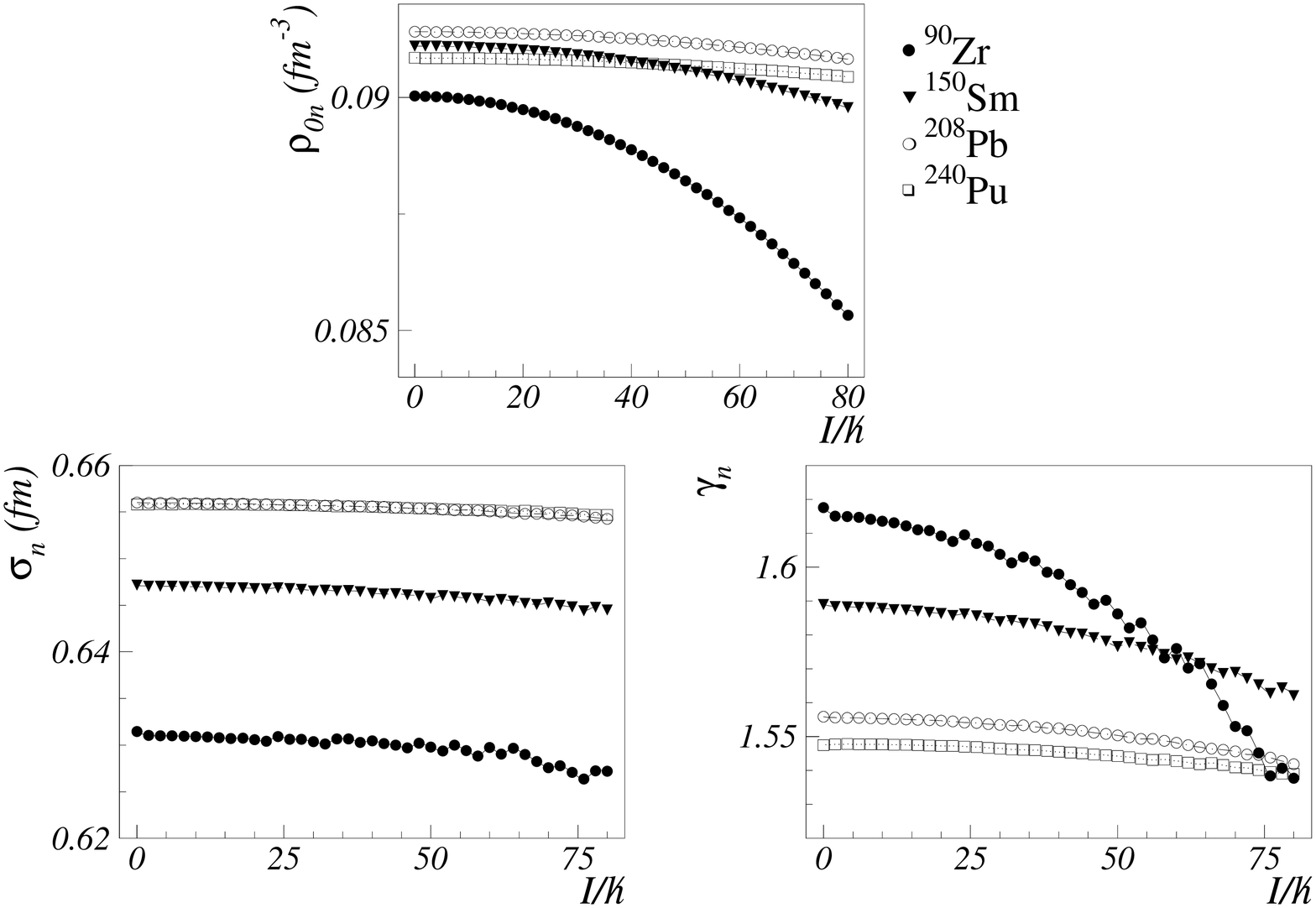}}
\bigskip
\caption{Variation with the angular momentum $I$ of three density parameters
($\rho_0$, $\sigma$, $\gamma$) defined in the text, for the neutron
distributions corresponding to four nuclei (\nuc{90}{Zr}, \nuc{150}{Sm},
\nuc{208}{Pb}, \nuc{240}{Pu}). The parameters $\sigma$ and $\rho_0$ are
expressed in fm and fm$^{-3}$ respectively.}
\bigskip
\label{nupar}\end{figure}

\begin{figure}[htb]
\centerline{\includegraphics*[width=.8\textwidth]{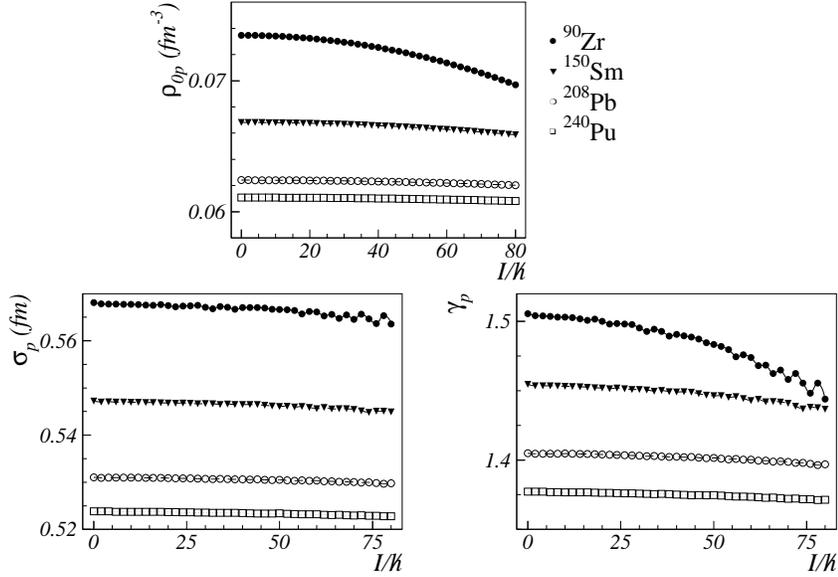}}
\bigskip
\caption{Same as Fig. \ref{nupar} for the proton distributions.}
\bigskip
\label{pipar}\end{figure}

\begin{figure}[htb]
\centerline{\includegraphics*[width=.8\textwidth]{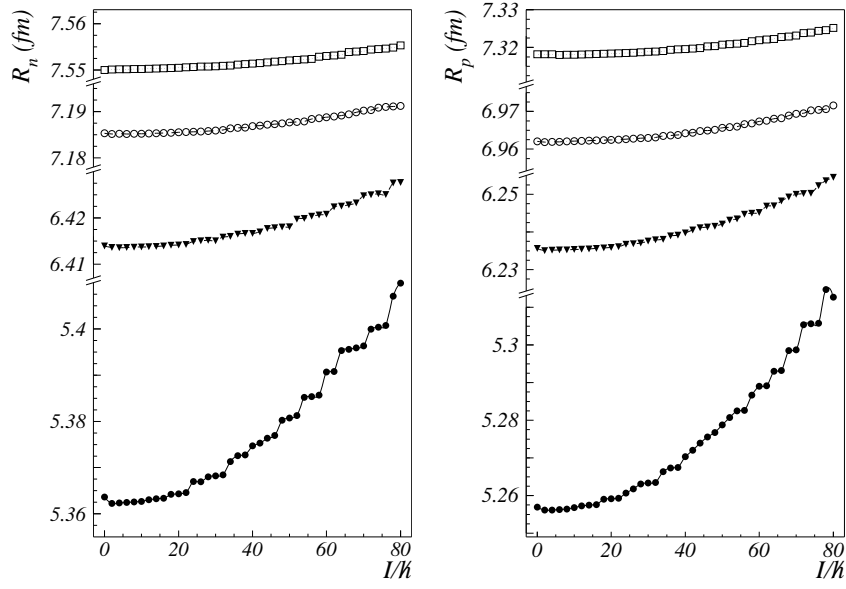}}
\bigskip
\caption{Variation with the angular momentum $I$ of the neutron ($R_n$) and
proton $(R_p$) sharp density radius parameters (expressed in fm) corresponding
to four nuclei (\nuc{90}{Zr}, \nuc{150}{Sm}, \nuc{208}{Pb}, \nuc{240}{Pu}).}
\bigskip
\label{radius}\end{figure}

The optimization procedure of the density has also been checked. The
termination criterion is delicate in any multidimensional minimization routine
and in the case of the density of Eq. (\ref{anisotropic}) we have 8
independent parameters to be optimized. In practice we use a downhill simplex
method (see e.g. \cite{NUM10}) and the numerical code stops when the decrease
in the function value in the terminating step is fractionally smaller than
some tolerance.  In our calculation this tolerance was set to 10$^{-7}$. Such
criteria might be fouled by a single anomalous step. In our case however, we
may use the fact that the calculated ETF energies are semi-classical and thus,
all the parameters are expected to vary slowly when changing the number of
particles, so that when the minimization is thoroughly performed for one
nucleus, we know that the deviation of the results for a neighbouring nucleus
should be rather small. The same is expected when going from a given value of
the angular momentum to a neighbouring one (i.e. here, differing by
2 $\hbar$). Therefore to make sure that we have reached a good local minimum,
the calculations have been run using as initial values of the parameters at a
given angular momentum, the optimized parameters as obtained at the preceding
value of the angular momentum (i.e. in our case lower by 2 $\hbar$).  To
illustrate this smooth behaviour, we display on Figs. \ref{nupar}-\ref{radius},
the angular momentum dependence of the density parameters for four nuclei
\nuc{90}{Zr}, \nuc{150}{Sm}, \nuc{208}{Pb} and \nuc{240}{Pu}. For the three
last nuclei, one gets gently varying density parameters up to spin values of
the order of 60 $\hbar$. The overall quality of these parameters is slightly
worse for \nuc{90}{Zr}.

Let us go back to the tolerance value adopted for the termination of the 
density parameters optimization. Such a level of relative accuracy has been 
found necessary in order to ensure the stability of the optimization process. 
However, it has been above noted that for light nuclei the computation of the
integral with respect to the cut-off radius was one order of magnitude
larger. In doing the optimization we must therefore assume that the latter
numerical error in evaluating the energy will not affect significantly the
localization of the optimal parameters. This has been shown to be true in a
limited number of test cases at zero spin.

Let us now briefly describe the fitting procedure in use here for the
energy. Mathematically, one minimizes for each value of the angular momentum
$I\hbar $ the weighted least-squares function
\begin{equation}
\chi^2(I) = \sum_{j=1}^{N_A} w_j \big[E_j^\mathrm{calc}(I)-E_j(a_k(I))\big]^2 ,
\end{equation}
with $E_j^\mathrm{calc}(I)$ being the nuclear part of the ETF calculated
energies or the Coulomb direct/exchange energies, $E_j(a_k(I))$ the
corresponding expression with $p$ parameters $a_k(I)$ to which we want to fit
the calculated energies and the weights $w_j$ have all been set to 1. As
before, $N_A$ is the number of nuclei used in the fit.

For this fit we used the code \textsc{afxy} \cite{AKD} which is an interactive
numerical researcher of nonlinear equations and nonlinear least square fits.
This algorithm is based on a composition of autoregularized Gauss-Newton
methods with improved versions of the Levenberg-Marquardt method and the
Tikhonov-Glasco descent-on-parameter method \cite{NUM15}. However, our fitting
problem is rather simple and it appears that a Gauss-Newton iteration process
is sufficient. The resulting root mean square deviation between calculated and
fitted energies, is defined as
\begin{equation}
\Delta E = \frac{1}{\sqrt{N_A}} \sqrt{\sum_{j=1}^{N_A} \big[E_j^{calc}(I) -
E_j(a_k^{opt}(I))\big]^2} ,
\end{equation}
where $a_k^\mathrm{opt}(I)$ are the optimized parameters (i.e. obtained upon
minimizing the $\chi ^2(I)$). Standard deviations of the parameters have been
obtained by usual numerical methods. The corresponding errors are listed in
Table \ref{fit-nuc}, together with the corresponding fitted values, for the
four parameters $(a_v,k_v,a_s,k_s)$ of $E_\mathrm{nuc}$ at four different
values of the angular momentum. As a result it appears that the significance
of the fitted values is much lower above $I=40$ $\hbar$, while it
remains qualitatively comparable below this value (we refer to the next
Section for a further discussion of the quality of the fit in terms of the
root mean square energy deviation).

\begin{table}[t]
\begin{center}
\caption{Fitted LDM-parameters and estimated errors (absolute values) for
different spin values.}
\label{fit-nuc}
\bigskip
\begin{tabular}{l|ll|ll|ll|ll}
$I/\hbar $ & \multicolumn{2}{|c}{0}  & \multicolumn{2}{|c}{20} 
           & \multicolumn{2}{|c}{40} & \multicolumn{2}{|c}{60} \\ \hline
& fit & error & fit & error & fit & error & fit & error \\ \hline
$a_v$ & -15.453 & 0.007 & -15.444 & 0.007 & -15.436 & 0.007 & -15.472 &0.013 \\
$k_v$ & 1.748 & 0.009 & 1.743 & 0.008 & 1.743 & 0.008 & 1.775 & 0.016 \\ \hline
$a_s$ & 16.961 & 0.037 & 16.908 & 0.036 & 16.848 & 0.037 & 17.016 & 0.069 \\
$k_s$ & 2.020 & 0.041 & 1.999 & 0.040 & 1.996 & 0.041 & 2.144 & 0.075 \\ \hline
\end{tabular}
\bigskip
\end{center}
\end{table}

\section{Results and discussion}

At first, we discuss the usefulness of the degree of freedom introduced in the
density to allow for a spatial anisotropy with respect to the angular velocity
[see Eq. (\ref{anisotropic})]. As a matter of fact, upon performing two series
of calculations (constraining the $\alpha_q$ parameters to vanishing values or
letting them free in the minimization procedure) we found, up to rather
high values of $\omega$ (up to $\hbar\omega=1.25$ MeV for a spherically
constrained \nuc{208}{Pb} nucleus corresponding to $I\simeq130$ $\hbar$), that
the gain in energy due to this degree of freedom was small
($\simeq110$ keV) and constant (up to a couple of keV). We therefore chose to
ignore this effect in what follows (i.e. set $\alpha_q=0$).

We will first consider the strong interaction part of the nuclear energy. Let
us assess the quality of our fit in terms of the root mean square energy
deviation $\Delta E$. For that we will study $\Delta E$ as a function of a
lower bound value $A_{\min}$ of the nucleon number of the retained nuclei
($A\geq A_{\min}$). The corresponding number of nucleons are reported in Table
\ref{cutoff}.

\begin{figure}[htb]
\centerline{\includegraphics*[width=.8\textwidth]{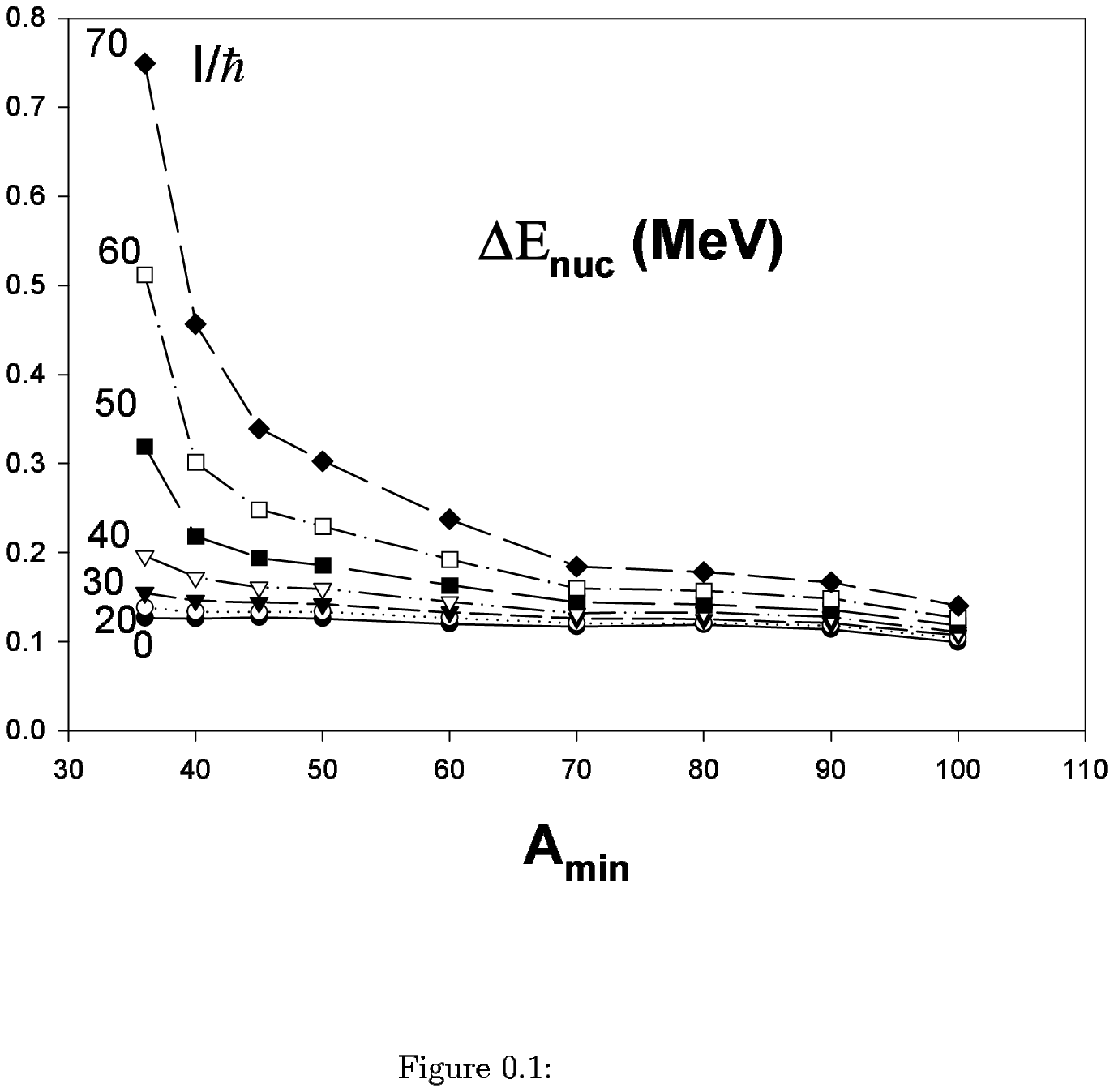}}
\bigskip
\caption{Root mean square energy deviation $\Delta E_{nuc}$ (in MeV) resulting
from the fits of the strong interaction part of the nuclear energy at various
values of the angular momentum $I$, as a function of the number of nuclei
$A_{min}$ defined in the text.}
\bigskip
\label{deltae}\end{figure}

The resulting energy deviations are plotted on figure \ref{deltae}. As a
result, it appears that for angular momenta which are not too high (lower than
30-40 $\hbar$) and for intermediate values of $A_{\min}$ (in the 60-80 range)
the quality of the fit seems rather stable. For large values of $A_{\min}$ the
size of the data basis is probably too limited. On the other hand light nuclei
at finite angular momenta $I$ seem to be not adequately represented by a
liquid drop formula (as seen on Fig. \ref{deltae}). This is more and more so
upon increasing $I$.

\begin{table}[t]
\begin{center}
\caption{Number of nuclei considered in the fit for a given lower bound
$A_{\min}$ of the total nucleon number $N_\mathrm{nucl}$.}
\label{cutoff}
\bigskip
\begin{tabular}{l|r|r|r|r|r|r|r|r|r|r}
$A_{\min}$        & 34 & 40 & 50 & 60 & 70 & 80 & 90 & 100 & 110 \\ \hline
$N_\mathrm{nucl}$ &212 &209 &204 &197 &181 &177 &170 & 157 & 146 \\ 
\end{tabular}
\bigskip
\end{center}
\end{table}

\begin{figure}[htb]
\centerline{\includegraphics*[width=.8\textwidth]{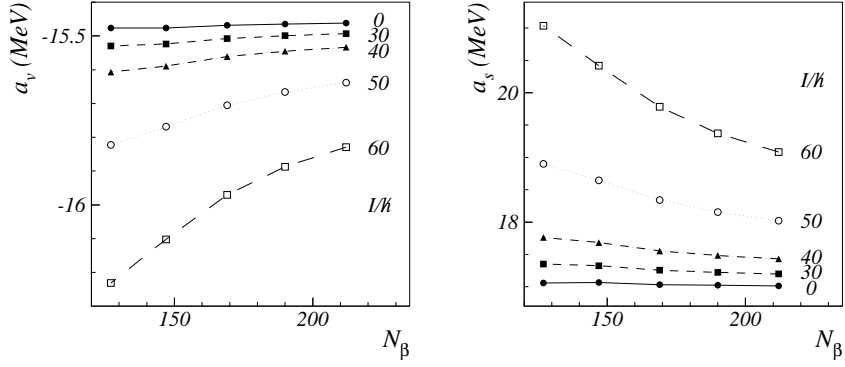}}
\bigskip
\caption{Fitted values of the usual volume ($a_v$) and surface ($a_s$) energy
coefficients expressed in MeV, as functions of the number of nuclei
$N_{\beta}$ defined in the text for various values of the angular momentum
$I$.}
\bigskip
\label{avas}\end{figure}

\begin{figure}[htb]
\centerline{\includegraphics*[width=.8\textwidth]{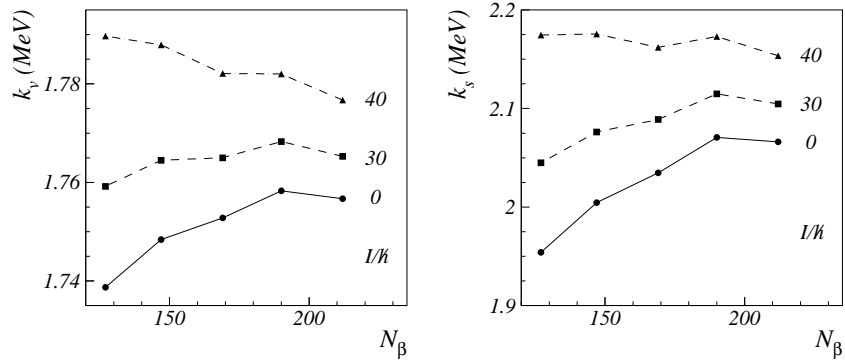}}
\bigskip
\caption{Same as Fig. \ref{avas} for the usual volume ($k_v$) and surface
($k_s$) asymmetry energy coefficients.}
\bigskip
\label{kvks}\end{figure}

As another test of the consistency of our data basis and of its physical
limitations, we have studied the convergence of the fitted LDM parameters $a_v$
and $a_s$ (Fig. \ref{avas}) as well as $k_v$ and $k_s$ (Fig. \ref{kvks}) in
terms of the number of considered nuclei $N_\beta$ varying as described
below. We define an approximate valley of stability as a function $N(Z)$. The
minimal distance of each considered nuclei to this valley is then computed.
Increasing $N_\beta$ corresponds to including into the fit nuclei which are
more and more distant from the stability curve.

The results displayed on Fig. \ref{avas} clearly demonstrate that the data
basis size is just barely adequate for rough estimates of the dependence of
the $a_v$ and $a_s$ coefficients as functions of $I$ at least for moderate
spin values ($I<30$-40 $\hbar$). As seen on Fig. \ref{kvks}, at the present
stage, the number of retained nuclei does not allow to draw safe quantitative 
conclusions on the $I$-dependence of the asymmetry coefficients $k_v$ and
$k_s$.

\begin{figure}[htb]
\centerline{\includegraphics*[width=.8\textwidth]{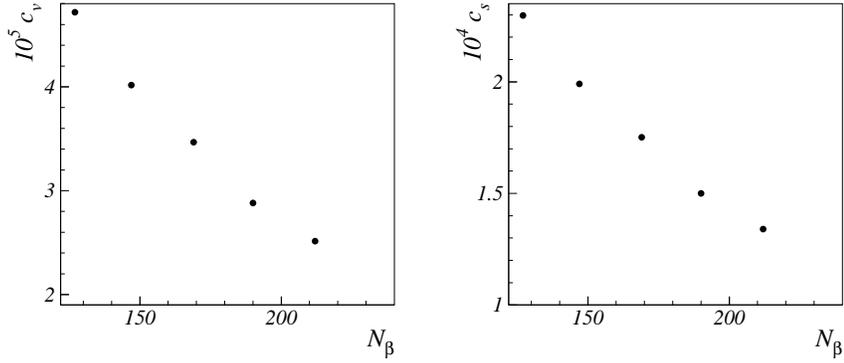}}
\bigskip
\caption{Coefficients $c_v$ and $c_s$ defining the angular momentum dependence
of the usual volume ($a_v$) and surface ($a_s$) energy coefficients as
functions of the number of nuclei $N_{\beta}$ defined in the text.}
\bigskip
\label{cvcs}\end{figure}

The results obtained on Fig. \ref{avas} are consistent with a quadratic
dependence of the $a_v$ and $a_s$ parameters as
\begin{equation}
a_v(I) = a_v(0) \left(1+ c_v I^2\right)
\end{equation}
and
\begin{equation}
a_s(I) = a_s(0) \left(1+ c_s I^2\right) .
\end{equation}
The results of the fit of $c_v$ and $c_s$, upon using all $a_v(I)$ and $a_s(I)$
calculated for $I$ going from 0 to 40$\hbar$ by 2 $\hbar$-steps, are plotted on
Fig. \ref{cvcs} as functions of the number of considered nuclei $N_\beta$ 
(defined as above described). These parameters are not converged, yet up to
about 10 percent one may assign to them the following ``asymptotic'' values
\begin{equation}
c_v \approx 2.0 \; 10^{-5} \qquad\text{and}\qquad c_s \approx 1.1 \; 10^{-4} .
\end{equation}

\begin{figure}[htb]
\bigskip
\centerline{\includegraphics*[width=.8\textwidth]{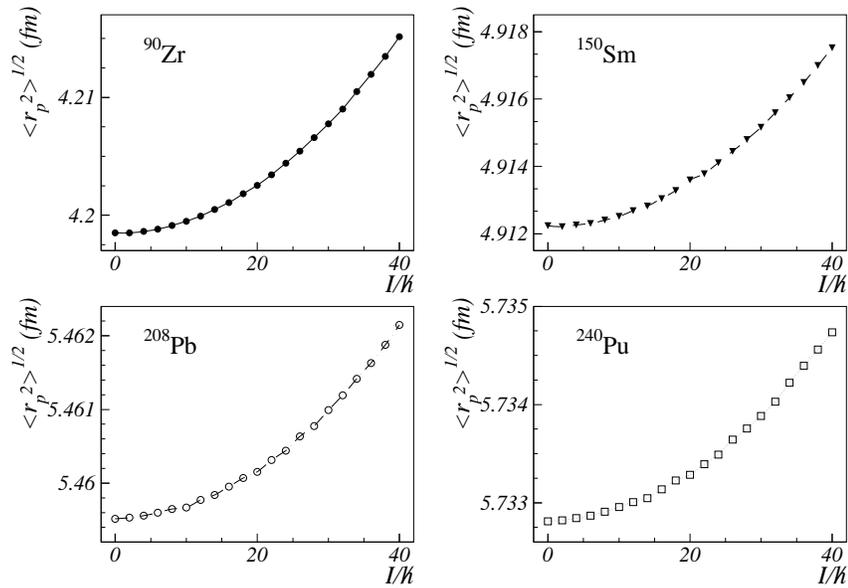}}
\caption{Root mean square radii of the proton distributions of four nuclei
(\nuc{90}{Zr}, \nuc{150}{Sm}, \nuc{208}{Pb}, \nuc{240}{Pu}) expressed in fm as
functions of the angular momentum $I$.}
\bigskip
\label{rms}\end{figure}

\begin{figure}[htb]
\centerline{\includegraphics*[width=.8\textwidth]{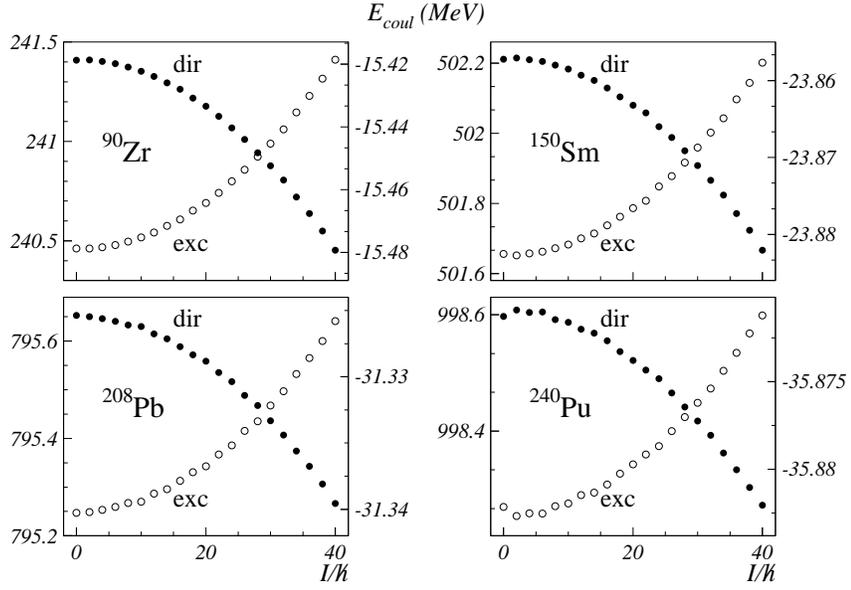}}
\bigskip
\caption{Direct (``dir'') and exchange (``exc'') Coulomb energies of four
nuclei (\nuc{90}{Zr}, \nuc{150}{Sm}, \nuc{208}{Pb}, \nuc{240}{Pu}) as
functions of the angular momentum $I$. The l.h.s. (r.h.s. resp.) scale
corresponds to the direct (exchange resp.) energies expressed in MeV.}
\bigskip
\label{coulene}\end{figure}

Let us now discuss the variation with angular momentum of the Coulomb
energies.  We present in Fig. \ref{rms} the evolution of the root mean square
radius of the proton density as a function of the angular momentum for the
four already considered nuclei. Due to the effect of centrifugal forces, it is
expected that the radius should increase with an increasing angular momentum
which is indeed the case here. As a consequence of their inverse proton radius
dependence, the Coulomb energies (both direct and exchange terms) plotted in
Fig. \ref{coulene} exhibit decreasing patterns in absolute values as functions
of $I$.

\begin{figure}[htb]
\centerline{\includegraphics*[width=.8\textwidth]{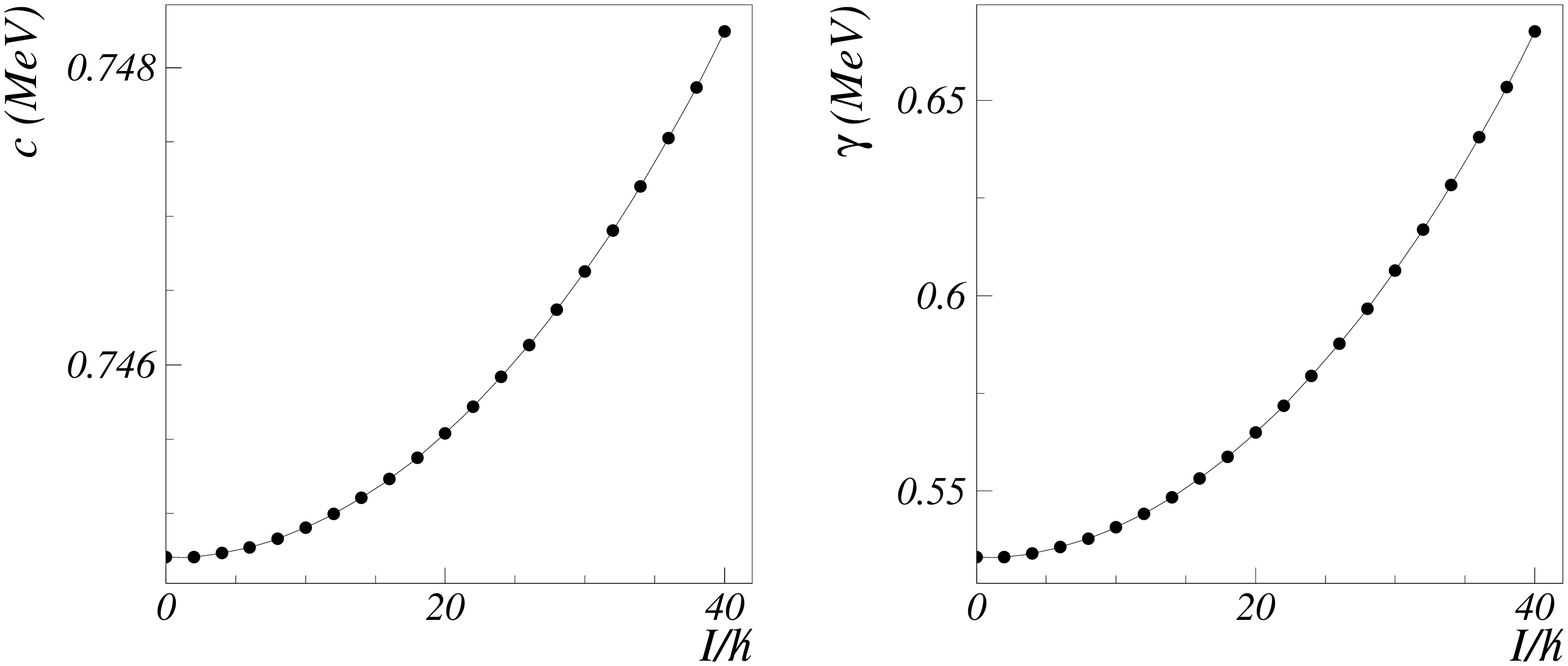}}
\bigskip
\caption{Dependence of the results of the 2-parameters fit of the direct part
of the liquid drop Coulomb energy in terms of the angular momentum $I$. Both
parameters $c$ and $\gamma$, defined in the text, are expressed in MeV.}
\bigskip
\label{couldir1}\end{figure}

\begin{figure}[htb]
\centerline{\includegraphics*[width=.8\textwidth]{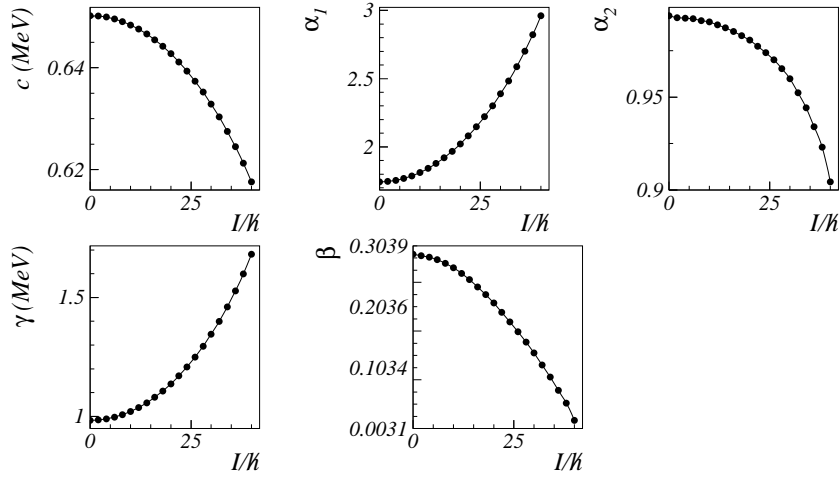}}
\bigskip
\caption{ same as Fig. \ref{couldir1} for the 5-parameters fit.}
\bigskip
\label{couldir2}\end{figure}

We have then performed a fit of the direct Coulomb energy $E_\mathrm{CD}$
using the two-parameter formula of Eq. (\ref{Ecdir1}). Since the Coulomb
energy is a decreasing function of $I$, one may have expected also to observe
a decreasing value of the parameter $c$ corresponding to the leading
contribution to $E_\mathrm{CD}$. As shown on Fig. \ref{couldir1} this does not
turn out to be the case. This incoherence is, in our opinion, due to the
inadequacy of Eq. (\ref{Ecdir1}) to describe consistently the isospin
dependence of the Coulomb energy.  As described in Ref. \cite{SBQ}, the
2-parameters formula of Eq. (\ref{Ecdir1}) provides a very poor fit of
$E_\mathrm{CD}$. Indeed it yields at zero spin for the same data basis
(i.e. including nuclei up to the drip-lines) a RMS energy deviation of some
MeV, which is close or above the maximum of the expected spin-dependence of
$E_\mathrm{CD}$. This is why we have performed the fit with the 5-parameters
formula of Eq. (\ref{Ecdir2}). With this new parameterization, we obtain
a parameter $c$ that now decreases with increasing $I$ (as shown on
Fig. \ref{couldir2} together with the variation of the other fitted
parameters).

\begin{figure}[htb]
\centerline{\includegraphics*[width=.8\textwidth]{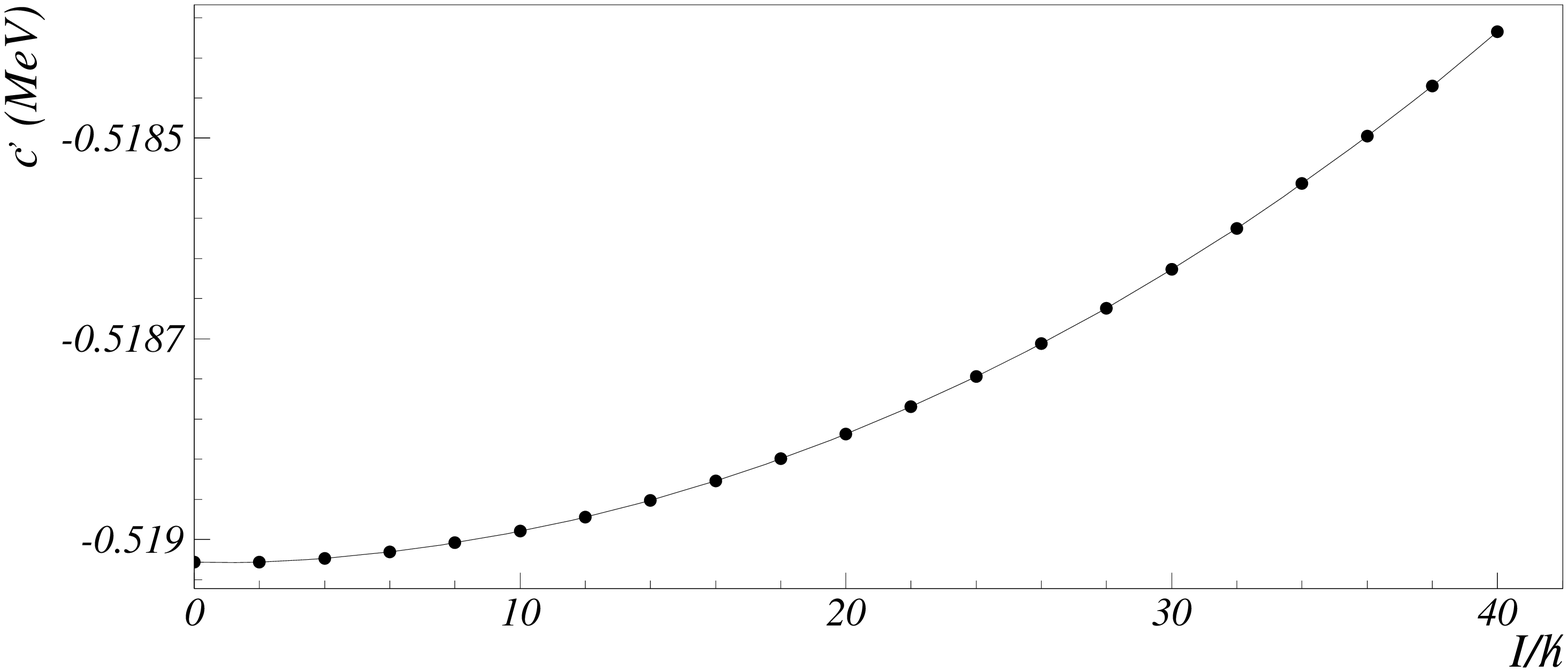}}
\bigskip
\caption{Dependence of the fitted value of the parameter $c'$, expressed in
MeV, of the liquid drop exchange part of the Coulomb energy defined in the
text, as a function of the angular momentum $I$.}
\bigskip
\label{coulexch}\end{figure}

As for the fit of the exchange term $E_\mathrm{CE}$, we obtain as demonstrated
on Fig. \ref{coulexch} a $c'$ parameter which decreases as a function of the
angular momentum. This behavior is consistent with the above described
variation of the proton radius.

\begin{figure}[htb]
\centerline{\includegraphics*[width=.8\textwidth]{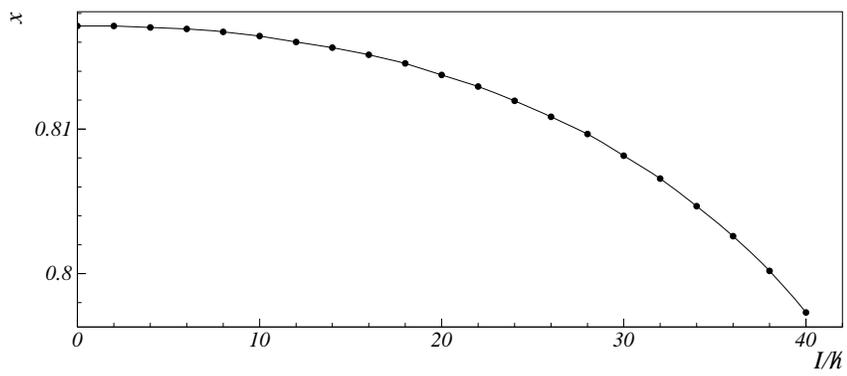}}
\bigskip
\caption{Usual fissility parameter $x$ of the \nuc{236}{U} nucleus as a
function of the angular momentum $I$.}
\bigskip
\label{fission}\end{figure}

The variations with $I$ of the parameters of both the strong interaction and
the Coulomb part of the liquid drop model energy entail some important
consequences on the nuclear deformability and hence on collective excitations
as well as on the fission stability. To illustrate the latter point we have
evaluated the fissility parameter $x$ defined as usual by
\begin{equation}
x=\frac{E_\mathrm{Coul}^0}{2E_s^0}
\end{equation}
(where $E_\mathrm{Coul}^0$ and $E_s^0$ stand for the Coulomb and surface
energies at zero deformation) of the \nuc{236}{U} nucleus for spin values
varying from 0 to 40 $\hbar$. In doing so, we took the surface energy as well
as Coulomb direct (5 parameters fit) and exchange energies resulting from our
fit.
It is to be noted that even though the variation with $I$ of the parameters
$k_s$ is not very well known quantitatively, the corrective character of the
asymmetry term with respect to the total surface energy ($\sim10$ percent)
allows us to draw anyway significant conclusions on the variation of $x$ with
$I$. As shown on Fig. \ref{fission}, the fissility parameter decreases by
$\sim2.5$ percent when increasing the angular momentum from 0 to
40 $\hbar$. According to the liquid drop estimate for the fission barrier
height $B_f$ of Ref. \cite{CS63}
\begin{equation}
B_f \simeq 0.83 \; E_s^0 \; (1-x)^3 \qquad\qquad\text{(for $x>2/3$)}
\end{equation}
One sees that such a variation of $x$ corresponds to an increase of $B_f$ by
approximately 30 percent for the considered \nuc{236}{U} nucleus. This result,
far from being insignificant, should be put into the perspective of the
rotating liquid drop approach of Ref. \cite{CPS82}. Upon increasing the
angular momentum, the rotational energy induces as well known an increased
instability against fission. The variation of the fissility parameter with the
spin which we have demonstrated here reduces the effective amount of the
resulting fission instability.

\section{Summary and conclusions}
In this work, we have attempted to estimate the angular momentum dependence
of the liquid drop model parameters and hence the validity of so-called
rotating liquid drop model approaches making use, as well known, of a
spin-independent liquid drop energy beyond the usual rotational energy
term. To do so, we have first performed for 212 different nuclei at 41
different angular momentum values, state of the art self-consistent
semi-classical calculations. Their results correspond to the $\hbar
\rightarrow 0$ solutions (beyond the trivial Thomas-Fermi approach) of a
time-reversal breaking Routhian problem using the full Skyrme SkM$^{*}$
effective force. A particular attention has been devoted to the numerical
stability of the corresponding variational solutions. The obtained strong
interaction and Coulomb energies have been separately fitted into liquid
drop model expressions at each angular momentum value. As a result, we have
shown that our theoretical data basis was just barely capable of providing
rough estimates of the angular momentum variations of the standard volume
($a_v$) and surface ($a_s$) parameters while providing only qualitative
insights into the behavior of the asymmetry parameters ($k_v$ and $k_s$).
The combined variation of the surface and Coulomb energies reflects itself
into a sizeable variation of the fissility parameter which yields, as a
consequence, an important enhancement of the fission stability for very
heavy nuclei, thus partially balancing the well-known instability generated
by the centrifugal forces. 

Clearly, one possible continuation of the present work should consist in
considerably enlarging the theoretical data basis in such a way as to extend
the level of confidence in the numerical estimates of the resulting $c_v$ and
$c_s$ parameters assessing the variations according to the angular momentum of
$a_v$ and $a_s$. Another goal should be to have also access to the
$I$-dependence of $k_v$ and $k_s$. The interest of such a numerical study
could be balanced by the fact that as soon as one is only interested in the
behavior of a few nuclei, one could as well perform ``exact'' self-consistent
semi-classical calculations. However it could be argued that such calculations
would necessitate, when performed for very deformed nuclei, or pieces of
nuclear matter exhibiting crevices or not simply connected, etc., highly non
trivial generalizations of the model densities in use for the variational
problem. At any rate, it remains that our present study clearly points out the
necessity to go beyond the use of a zero angular momentum parameter set when
dealing with a rotating liquid drop model approach.

Some other lines of further work could be mentioned. One would consist in
evaluating carefully and systematically the combined influence of the
rotational energy term and of the spin-dependent liquid drop energy term
on fission properties. The other set of fascinating questions stems from
the negative result obtained here when trying to identify an anisotropy
effect in the spatial matter distribution. It is indeed not clear to us
that such a negative result should survive whenever one would deal with
strongly deformed solutions. Moreover, the semi-classical solutions in use
here within the scheme proposed by Grammaticos and Voros \cite{GV80,GV79},
result from an angular averaging in the momentum space. Studying the
anisotropy in this momentum space might come out at large angular velocities
as another source, and may be the main source, of discrepancy between the
naive spin-independent liquid drop model approach and a fully self-consistent
microscopic approach of pieces of nuclear matter experiencing fast collective
rotations.

\section*{Acknowledgments}
During the course of this work, we have benefitted from numerous
discussions with many physicists among which we would like to particularly
thank G. Barreau, E. Chabanat, D. Karadjov and I.N. Mikha\"\i lov. This work
for a  part, has been sponsored through grants provided by the Bulgarian
Academy of Sciences/CNRS and DRS/CNRS (respectively Bulgaria/France
and Algeria/France) agreements which are gratefully acknowledged.

\vfill\eject

\vfill

\end{document}